
%
%
\magnification=1200
\hsize=4.5 in
\hoffset=1.0 cm
 \overfullrule = 0pt
\line{ }
\vskip 0.5 truecm
\rightline {DFTUZ-91.33}
\rightline {November 1991}
\vskip 2. truecm
\centerline{\bf RENORMALIZATION GROUP AND TRIVIALITY}
\vskip 0.15 truecm
\centerline{\bf IN NONCOMPACT LATTICE QED WITH LIGHT FERMIONS.}
\vskip 2 truecm
\centerline { V.~Azcoiti }
\vskip 0.15 truecm
\centerline {\it Departamento de F\'\i sica Te\'orica, Facultad de
Ciencias,
50009 Zaragoza (Spain)}
\vskip 0.5 truecm
\centerline { G. Di Carlo and A.F. Grillo }
\vskip 0.15 truecm
\centerline {\it Istituto Nazionale di Fisica Nucleare, Laboratori
Nazionali di Frascati,}
\centerline {\it P.O.B. 13 - Frascati (Italy). }
\vskip 3 truecm
\centerline {ABSTRACT}
\vskip 0.3 truecm
\par
{In the framework of noncompact lattice QED with light fermions, we
 derive the functional dependence of the average energy per plaquette
on the bare parameters using block-spin Renormalization Group
arguments and assuming that the renormalized coupling vanishes.
Our numerical results for this quantity
in $8^4$ and $10^4$ lattices show evidence
for triviality in the
weak coupling phase and point to a non vanishing value for the renormalized
coupling constant
in the strong coupling phase.}
\vskip4truecm
\vfill\eject
\par
One of the most succesfull results of lattice regularization of gauge theories,
combined with Monte Carlo simulations has been a deeper understanding of the
non perturbative dynamics of asymptotically free gauge theories, like QCD.
Conversely, very little is known about non asymptotically free theories.
\par
As a matter of fact, the most important question for lattice regularization,
i.e. the existence of a quantum continuum limit with non trivial dynamics, has
not (yet) be answered for the simplest and perturbativelly most succesfull
gauge theory, namely, Quantum Electrodynamics.
\par
In recent years, many efforts have been devoted to the study of this problem,
firstly using the compact regularization of the abelian model [1,2,3]
and more recently within the non compact formulation .
\par
The first numerical investigations of the non compact model, in the quenched
approximation [4], have shown the existence
of a continuos chiral transition at
finite value  of the coupling constant. This transition survives after the
inclusion of dynamical fermions [5-8] so suggesting that the quantum
continuum physics could be reached there.
\par
Having found a candidate point for the continuum limit, two important
questions should be answered:
i) Is the theory defined by taking the limit at the chiral
critical point non trivial, i.e. does this model possess a particle
spectrum with non trivial interactions in this limit? and
ii) Assuming answer to i) is positive, has this limiting theory
something to do with standard quantum electrodynamics?
\par
Concerning the first point, there exist extensive numerical simulations
performed by several groups [5-9,11-13]. The first indication of a power-law
(as opposed to essential singularity [10]) scaling for the chiral condensate
was suggested by A. Horowitz in [11]. On the other hand, the Illinois group
found a good quantitative support for a non mean-field power law scaling in
the quenched model [12,13]. Their results ruled out the mean field scenario and
illustrated the degree of difficulty required in extracting the critical
indices in the full theory with dynamical fermions, where larger lattices
and a precise determination of the critical coupling are necessary in order
to compute critical exponents [9].
\par
On the other hand, G\"ockeler et al. [14,15]
computed the renormalized
charge and fermion mass and found that the corresponding
Callan-Symanzik $\beta$ function is consistent with the prediction of
renormalized perturbation theory.
Furthermore,
they have not found lines of constant physics for the matter sector in the
two parameters region they explored [15]; on the basis of their results they
argue about the non-renormalizability of the theory.

\par
As for point ii) above, Hands et al. [9] have shown that the vacuum
in the broken phase of non compact QED is a monopole condensate with
U(1) symmetry while the continuum model has no finite action
monopoles, and the gauge group symmetry is $\Re$. This means that the
lattice model is qualitatively different from standard QED, belonging
to a different universality class. This result also casts serious
doubts on the validity of renormalization group flow calculations as
those of refs. [14,15].
\par
In this letter we report a study of the triviality problem of the quantum
continuum limit of non compact Lattice QED. To this end, we introduce a
new approach based on a characterization of the behaviour of the mean
plaquette energy as a funtion of the bare parameters $\beta$ and
$m$. The equation describing such a
behaviour holds only if the continuum limit of the model consists of
particles without electromagnetic interactions.
\par

Consider the action of non compact lattice QED

$$S = {1 \over 2}\sum_{x,\mu}\eta_\mu(x)
{\bar \chi}(x)\{U_\mu(x)\chi(x+\mu)-U_\mu^*(x-\mu)\chi(x-\mu)\}+$$
$$ m \sum_x{\bar \chi}(x)\chi (x) +{\beta\over 2} \sum_{x,\mu < \nu}
F_{\mu \nu}^2(x)\eqno(1)$$
$$F_{\mu \nu}(x)=
A_\mu(x) + A_\nu(x+ \hat \mu) - A_\mu(x+\hat \nu) - A_\nu(x) $$

\noindent
where $\beta=1/e^2$ and we use staggered fermions coupled to the gauge
fields $A_\mu(x)$ through the compact link variable $U_\mu(x)$.

A problem working in the non compact formulation comes from the fact that
the partition function associated to action (1) is not well defined
even in a lattice of finite size. In fact the gauge group integration,
in contrast to the compact case, is divergent.
The problem can be overcome by gauge fixing. We instead factorize the
divergency in the density of states as follows.

Define the density of states at fixed non compact normalized energy
$E$ in a lattice of volume $V$

$$ N(E) =\int [d{A_\mu(x)}] \delta({1\over 2} \sum_{x,\mu < \nu}
F_{\mu \nu}^2(x) - 6VE) \eqno(2)$$

\noindent
$N(E)$ is divergent because of the infinite volume of gauge
integration. However, this divergence can be factorized out and
one can easily show that

$$ N(E) = C_G (6VE)^{{3 \over 2} V - 1}   \eqno(3) $$

\noindent
where $C_G$ is a divergent constant (the volume of the gauge group).

On the other hand it can be shown, following ref. [3], that the partition
function can be written as an integral over the normalized non compact
energy $E$

$${\cal Z} = \int dE N(E) e^{-6 \beta V E} e^{-S_{eff}^F(E,m)} \eqno(4)$$

\noindent
where

$$  e^{-S_{eff}^F(E,m)} =  {\int [dA_{\mu}(x)] det
\Delta(m, A_{\mu}(x))
\delta({1\over 2} \sum_{x,\mu < \nu} F_{\mu \nu}^2(x) - 6VE)
\over
\int [dA_{\mu}(x)]
\delta({1\over 2} \sum_{x,\mu < \nu} F_{\mu \nu}^2(x) - 6VE)}
\eqno(5)$$

{}From eq. (3),(4),(5) we can derive an effective action for the full
theory in the thermodinamical limit $V \to \infty$ as

$$ S_{eff}(E,V,\beta,m) =
-{3\over 2} V ln E + 6 \beta V E + S_{eff}^F(E,m)  \eqno(6) $$

Now let us write the partition function associated to (1) as an integral over
the plaquette variables $F_{\mu \nu}^2$ in the following way

$$ Z =\int [d{A_\mu(n)}][d{\bar \chi}(n)][d{\chi (n)}] [dE_{\mu \nu}(n)]
\prod\delta(F_{\mu \nu}^2(n) - E_{\mu \nu}(n)) e^{-S} = $$
$$ \int [dE_{\mu \nu}(n)] N(E_{\mu \nu}(n)) e^{-S(E_{\mu \nu}(n))}
\eqno(7)$$

\noindent
where

$$  e^{-S(E_{\mu \nu}(n))} = $$

$$ {\int [dA_{\mu}(n)]
[d{\bar \chi}(n)][d\chi (n)]\prod
\delta(F_{\mu \nu}^2(n) - E_{\mu \nu}(n)) e^{-S}}
\over
\int [dA_{\mu}(n)]
[d{\bar \chi}(n)][d\chi (n)]\prod
\delta(F_{\mu \nu}^2(n) - E_{\mu \nu}(n))  \eqno(8)$$

\noindent
and the denominator in (8) is just the density of states $N(E_{\mu \nu}(n))$.

Next, imagine we apply linear block-spin renormalization group transformations
to the theory described by the effective action
$S(E_{\mu \nu}(n)) - ln N(E_{\mu \nu}(n))$ . Our spin variable is the plaquette
variable $E_{\mu \nu}(n)$ which takes values from $0$ to $\infty$
and blocking
is performed at each $\mu \nu$ plane. We generate in this way a
series of effective actions which are equivalent at large distances since
we are integrating out all the short distance details. If the theory is
trivial i.e., if all renormalized couplings vanish, the only relevant
parameter at the end of this procedure will be the coefficient of the kinetic
term $E_{\mu \nu}(n)$. Then, the renormalized action
$S_{R}(E_{\mu \nu}(n))$ will be, apart from the density of states
contribution, of the form

$$ S_{R}(E_{\mu \nu}(n)) =
{1\over 2} \bar \beta(m,\beta)\sum_{n,\mu < \nu} E_{\mu \nu}(n) +
 h(m,\beta)
\eqno(9)$$

\noindent
where $\bar \beta(m,\beta)$ and $h(m,\beta)$ are unknown renormalized
constants.
Defining  $E_R = {1\over 6V} \sum_{n,\mu < \nu} E_{\mu \nu}(n)$ we get

$$ S_R^{eff}(E_R) =
-{3\over 2} V ln E_R + 6 \bar \beta V E_R +
h(m,\beta)
\eqno(10) $$

This action and action (6) can differ only by a multiplicative factor
$X(m,\beta)$ in the mean energy $E$ since we have obtained (10) by means of
linear block-spin transformations plus a final linear global transformation.
Therefore triviality means that action (6),
apart from the logarithmic term coming from the density of states, must be
a linear function of the mean energy $E$ or equivalently, that
$S_{eff}^F(E,m)$ in (6) is a
linear function of $E$.

We would like to remark at this point that the connection between actions
(6) and (10) can be established owing to the use of linear block-spin
Renormalization Group transformations, so that we can obtain eq. (10) from
(6) through a linear change of variables.
Using non linear transformations or
transformations in other kind of variables, we could identify the partition
functions but we would not be able to establish any connection between the
corresponding effective actions.

Due to the fact that $S_{eff}^F(E,m)$ is a linear
function of E, we get that all effects of dynamical
fermions can be reduced to a redefinition of the coupling constant $\beta$.
Therefore, the mean plaquette energy can be written as

$$E(m,\beta) = {1 \over 4(\beta+h_1(m))} \eqno(11) $$

The linearity of the effective action (6) as well as equation (11), which
should hold around the critical point if the theory is trivial, can be
compared with data obtained by numerical simulations.

Following a method that we have recently proposed [3], we
calculated the mean plaquette energy using the fermionic
effective action (5).
We have obtained the fermionic effective action in 27 values of
$E$ in the range $0.5 - 1.7$, allowing us to calculate
thermodynamical quantities as a function of $\beta$ in the range
$0.14 \le \beta \le 0.40$; with these values we go deeply inside
the strong coupling and Coulomb phase respectively.

The largest part of the simulations has been performed on a $8^4$
lattice, but from an analysis of the scaling properties of fermionic
effective action in lattices from $4^4$ to $10^4$, we can exclude
significant finite volume effects on the mean plaquette energy of the
full theory, already in the $8^4$ lattice. For a detailed report of these
simulations see [16].

In Fig. 1 we report the effective fermionic action (5) for vanishing
fermion mass as a function of $E$. Two different regimes,
corresponding to two different phases, can be seen from this
figure.

Coulomb phase ($\beta>0.206$) which is dominated in the thermodynamical limit
by energies $(E \leq 1.0)$, is characterized by an
effective action linear in $E$, meaning that the effect of the
inclusion of fermionic degrees of freedom merely reduces to a shift in
the coupling constant, indicating triviality. In fact, if we try
to fit the plaquette energy data in this phase
with a functional form like (11), we
obtain a very good fit for $h_1=0.0409$ ( see Fig. 2 ). The fact that
(11) is able to reproduce in such a good way the numerical data is
again a strong indication of triviality in this phase.

Completely different is the situation in the strong coupling broken
phase $(E > 1.0)$. Indeed, the
behavior of the plaquette energy for $\beta<0.206$ deviates from the fit (11),
this indicating the existence of a phase transition at $\beta_c\simeq 0.206 $.
For $\beta<\beta_c$ we tried to fit our data
with a function like (11). However, we have found
that we need to give a $\beta$ dependence to $h_1(m)$, as can
be seen in Fig. 3. From the two fits for $h_1(m)$ in this Figure, we get
$\beta_c = 0.206(5) $.

Our results on the effective fermionic action reported in Fig.1 can be very
well understood if a second order phase transition occurs.
Indeed, applying the saddle point technique to the computation of the partition
function (4) it can be shown that a discontinuity in the
specific heat implies a discontinuity of the second energy derivative of the
effective fermionic action at the  energy critical value. Furthermore
and as following from the main content of this paper, a non vanishing value for
the renormalized coupling is directly related to a non linear energy
dependence of the effective fermionic action. Therefore a second order phase
transition should produce a discontinuity of the renormalized coupling at the
critical point.

Fitting the points in Fig.1 by two polynomials, one for $E<1.018$
and the other for $E>1.018$ (continuous line in the Figure), being
$E_c = 1.018$ the mean energy at $\beta = 0.206$ $m = 0$, we get very
good fits with a gap of $0.38(2)$ in the second energy derivative
normalized by the lattice volume . As a result of the
fits we also find that the first energy derivative of the efective
fermionic action is continuous
at $E = E_c$ and second and higher order energy derivatives vanish for
$E < E_c$ inside the errors of the fits, these results being very stable when
we encrease the degree of the polynomial fits.
The observed approximate scaling of
the effective fermionic action with the lattice volume when we go from the
$6^4$ to the $10^4$ lattice [16] implies that finite size effects does not
affect our results in a significant way. In any case, the important
qualitative finding is that the second energy derivative of effective
fermionic action is always discontinuous at the critical energy.

In conclusion, our numerical analysis shows the existence of a phase transition
at $\beta_c = 0.206(5)$, $N_f = 4$, $\beta_c = 0.226(5)$, $N_f = 2$,
in agreement within  errors with the critical value obtained
from the behaviour of the chiral condensate [9, 16].
The behavior of the effective fermionic
action and mean plaquette energy in the broken phase, strongly suggests
a non vanishing value for the renormalized coupling constant in this phase,
even when we approach the critical point.

Does this result implies the existence of a non trivial (non gaussian) fixed
point?. In the general formulation of the Renormalization Group approach it is
generally assumed that any point at or near the critical surface is in the
attraction domain of some fixed point, even though singular
behaviour can not be excluded by general arguments [17]. Excluding
such a singular behavior, our numerical results strongly indicate that the
fixed point is non gaussian.

The important question now is: is the quantum theory described by this fixed
point renormalizable?.
The results reported in [15] about non perturbative
renormalizability of the model show that there are no lines of constant
physics in the ($\beta$, m) plane. However, this result does not imply
necessarily non renormalizability since it could be that the two
dimensional parameter space is too small. In fact, in the two parameters
action (1) we have neglected coupling terms such as four Fermi interactions
and monopole contributions which can be generated in the renormalization
procedure and whose associated couplings could become relevant for the
continuum limit in the strong coupling phase, as suggested firstly in [18]
and also by the results of ref. [9] (this was also the possibility left open
in [15]). If this is the real scenario, our numerical results in the broken
phase should be regarded as a strong indication for a non trivial continuum
limit.

On the contrary, when the transition is approached from the Coulomb phase,
as more appropriate for the definition of continuum QED, the theory is non
interacting. This behaviour is not totally unexpected, since
we know from perturbative QED the existence of Landau pole problem.

The authors aknowledge J.L. Alonso and J.L. Cortes for useful discussions.
This work has been partly supported through a CICYT (Spain) -
INFN (Italy)
collaboration.

\vfill
\eject

\line{}
\vskip 1 truecm
\centerline {\bf REFERENCES}
\vskip 1 truecm
\item {1.}{V. Azcoiti, A. Cruz, E. Dagotto, A. Moreo and A. Lugo,
 Phys. Lett. {\bf 175B} 202 (1986).\hfill}
\vskip .1 truecm
\item {2.} {J.B. Kogut and E. Dagotto,
 Phys. Rev. Lett. {\bf 59} 617 (1987);
 E. Dagotto and J.B. Kogut,
 Nucl. Phys. {\bf B295[FS2]} 123 (1988).\hfill}
\vskip .1 truecm
\item {3.}
{V. Azcoiti, G. Di Carlo and A.F. Grillo,
 Phys. Rev. Lett. {\bf 65} 2239 (1990);
 V. Azcoiti, A. Cruz, G. Di Carlo, A.F. Grillo and A. Vladikas,
 Phys. Rev. {\bf D43} 3487 (1991).\hfill}
\vskip .1 truecm
\item {4.} {J. Bartholomew, S.H. Shenker, J. Sloan, J.B. Kogut, M. Stone,
 H.W. Wyld, J. Shigmetsu and D.K. Sinclair,
 Nucl. Phys. {\bf B230} 222 (1984).\hfill}
\vskip .1 truecm
\item {5.} {J.B. Kogut, E. Dagotto and A. Kocic,
 Phys. Rev. Lett. {\bf 60} 772 (1988);
 Nucl. Phys. {\bf B317} 253 (1989);
 Nucl. Phys. {\bf B317} 271 (1989).\hfill}
\vskip .1 truecm
\item {6.} {E. Dagotto, A. Kocic and J.B. Kogut,
 Nucl. Phys. {\bf B331} 500 (1990).\hfill}
\vskip .1 truecm
\item {7.}{S.P. Booth, R.D. Kenway and B.J. Pendleton,
 Phys. Lett. {\bf 228B} 115 (1989).\hfill}
\vskip .1 truecm
\item {8.} {M. G\"ockeler, R. Horsley, E. Laermann, P. Rakow, G. Schierholz,
 R. Sommer and U.J. Wiese,
 Nucl. Phys. {\bf B334} 527 (1990).\hfill}
\vskip .1 truecm
\item {9.} {S.J. Hands, J.B. Kogut, R. Renken, A. Kocic, D.K. Sinclair
 and K.C. Wang,
 Phys. Lett. {\bf 261B} 294 (1991).\hfill}
\vskip .1 truecm
\item {10.} {V. Miransky,
 Nuovo Cim. {\bf A90} 149 (1985).\hfill}
\vskip .1 truecm
\item {11.} {A.M. Horowitz,
 Nucl. Phys. {\bf B17} (Proc. Suppl.) 694 (1990);
 Phys. Lett. {\bf 244B} 306 (1990).\hfill}
\vskip .1 truecm
\item {12.} {A. Kocic, S. Hands, J.B. Kogut and E. Dagotto,
 Nucl. Phys. {\bf B347} 217 (1990).\hfill}
\vskip .1 truecm
\item {13.} {E. Dagotto, J.B. Kogut and A. Kocic,
 Phys. Rev. {\bf D43} R1763 (1991).\hfill}
\vskip .1 truecm
\item {14.} {M. G\"ockeler, R. Horsley, E. Laermann, P. Rakow, G. Schierholz,
 R. Sommer and U.J. Wiese,
 Phys. Lett. {\bf 251B} 567 (1990).\hfill}
\vskip .1 truecm
\item {15.} {M. G\"ockeler, R. Horsley, P. Rakow, G. Schierholz and R. Sommer,
 DESY preprint {\bf DESY 91-098} (1991).\hfill}
\vskip .1 truecm
\item {16.} {V. Azcoiti et al. ,
 DFTUZ preprint {\bf DFTUZ 91.34} (1991).\hfill}
\vskip .1 truecm
\item {17.} {K.G. Wilson and J.B. Kogut,
 Phys. Rep. {\bf 12C} 75 (1974).\hfill}
\vskip .1 truecm
\item {18.} {W.A. Bardeen, C.N. Leung and S.T. Love,
 Nucl. Phys. {\bf B273} 649 (1986);
 Nucl. Phys. {\bf B323} 493 (1989).\hfill}

\vfill
\eject
\line{}
\vskip 1 truecm
\centerline{\bf FIGURE CAPTIONS}
\vskip 1 truecm
\item {1)}{ Effective fermionic action (5), versus $E$ at $m=0.0$ and four
flavors,
obtained through microcanonical simulation.
Statistical errors are invisible at this scale. }

\vskip .1 truecm

\item {2)}{ Mean plaquette energy $E(m=0,\beta)$ versus $\beta$.
Solid line is a fit, equation (11), with $h_1=0.0409$. Errors are
of the order of symbols size. }

\vskip .1 truecm

\item {3)}{$h_1(m)$ versus $\beta$ at $m=0.0$. In the weak coupling phase
($\beta > 0.206$), $h_1(m)$ is well fitted by a horizontal line.
For $\beta<\beta_c$ equation (11) does not hold and we need to give a $\beta$
dependence to $h_1(m)$. The solid line in the strong coupling phase is a
polynomial fit. }

\vfill
\eject
\end